\documentclass[preprint]{aastex631}

\usepackage{hologo} % for TeX engine logos
\usepackage{booktabs}
\usepackage{multirow}
\usepackage{makecell}
\usepackage{xcolor}
\usepackage{ulem}
\usepackage{amsmath}

\graphicspath{{./}{figures/}}

\usepackage{graphicx} % Required for including images
\usepackage{float} % Required for [H] placement specifier

\usepackage{CJK}

\begin{document}
\begin{CJK*}{UTF8}{gbsn}

\title{Reconstruction of the dark energy scalar field potential by Gaussian process}

\author[0000-0002-6495-0158]{Jing Niu (牛菁)}
\affiliation{Institute for Frontiers in Astronomy and Astrophysics, Beijing Normal University, Beijing 102206, China}
\affiliation{School of Physics and Astronomy, Beijing Normal University, Beijing 100875, China}

\author[0000-0003-0167-9345]{Kang Jiao (焦康)}
\affiliation{School of Physics and Laboratory of Zhongyuan Light, Zhengzhou University, Zhengzhou 450001, China}

\author{Peng He (贺鹏)}
\affiliation{Bureau of Frontier Sciences and Basic Research, Chinese Academy of Sciences, Beijing 100190, China}

\author[0000-0002-3363-9965]{Tong-Jie Zhang (张同杰)}
\affiliation{Institute for Frontiers in Astronomy and Astrophysics, Beijing Normal University, Beijing 102206, China}
\affiliation{School of Physics and Astronomy, Beijing Normal University, Beijing 100875, China}

\correspondingauthor{Tong-Jie Zhang}
\email{tjzhang@bnu.edu.cn}



\begin{abstract}

Dark energy is believed to be responsible for the acceleration of the universe. In this paper, we reconstruct the dark energy scalar field potential $V(\phi)$ using the Hubble parameter $H(z)$ through Gaussian Process analysis. Our goal is to investigate dark energy using various $H(z)$ datasets and priors. We find that the selection of prior and the $H(z)$ dataset significantly affects the reconstructed $V(\phi)$.  And we compare two models, Power Law and Free Field, to the reconstructed $V(\phi)$ by computing the reduced chi-square. The results suggest that the models are generally in agreement with the reconstructed potential within a $3\sigma$ confidence interval, except in the case of Observational $H(z)$ data (OHD) with the Planck 18 (P18) prior. Additionally, we simulate $H(z)$ data to measure the effect of increasing the number of data points on the accuracy of reconstructed $V(\phi)$. We find that doubling the number of $H(z)$ data points can improve the accuracy rate of reconstructed $V(\phi)$ by 5$\%$ to 30$\%$.

\end{abstract}

\keywords{Dark energy (351) --- Astronomy data analysis (1858)}


\section{Introduction} \label{sec:Introduction}

In 1998, two independent teams, \cite{1998AJ....116.1009R} and \cite{1999ApJ...517..565P}, discovered that the expansion of the universe is accelerating by observing the distant type  \uppercase\expandafter{\romannumeral1}a supernovae. Since then, cosmic acceleration has been accepted by cosmologists. Understanding the origin of acceleration is critical to understanding fundamental physics in the universe. There are two explanations for the cosmic acceleration \citep{2008ARA&A..46..385F}. One is dark energy \citep{2018RPPh...81a6901H}, which is a new energy form with large negative pressure having repulsive gravity. The other is modified gravity \citep{2019LRR....22....1I}, where general relativity breaks down on cosmological scales.

The simplest model for dark energy is vacuum energy, which mathematically corresponds to a cosmological constant that does not vary with space or time. In this model, dark energy is not a dynamic quantity. However, by introducing a new degree of freedom in the form of a scalar field $\phi$, dark energy can become dynamical \citep{1988NuPhB.302..668W,1988PhRvD..37.3406R,1995PhRvL..75.2077F,1999PhRvL..82..896Z}. Recently, the results from \cite{2024arXiv240403002D} have motivated the search for dynamical dark energy models. The measurements of baryon acoustic oscillations (BAO) from DESI, when combined with either the cosmic microwave background (CMB) or type \uppercase\expandafter{\romannumeral1}a supernovae individually, prefer $\omega_0 > -1$ and $\omega_a < 0$. There are also other dark energy models, including more complex scalar-field models \citep{2002PhLB..545...23C,2003PhRvD..68b3509C}. In this article, we investigate the scalar field using observational $H(z)$ data (OHD) \citep{2013PhRvD..88j3528Y} via Gaussian Process. OHD can be measured using different ways, such as cosmic chronometers (CC) and baryon acoustic oscillations (BAO). Unlike supernova observations, which measure $H(z)$ indirectly through its integral along the line of sight, these methods can directly measure $H(z)$ \citep{2018MNRAS.477.2867J}.

Quintessence is a type of scalar field first introduced by  \cite{1998PhRvL..80.1582C}. According to this theory, quintessence represents a form of dark energy that changes over time. In order to study the properties of dark energy, it is useful to study the scalar field $\phi$ endowed with a dark energy scalar field potential $V(\phi)$. Recently, several works have focused on reconstructing the dark energy scalar field potential from cosmological data sets without making any assumptions about its form. For example, \cite{2022JCAP...11..037J} used $H(z)$ and Type Ia supernovae data to constrain the dark energy potential via Gaussian Process. They also compared the results obtained using different priors, namely Planck 18 and a large prior to assess the differences in the reconstruction. \cite{2022arXiv220306767E} used 40 $H(z)$ data points to reconstruct the dark energy potential through Gaussian Process in a model-independent manner. They compared the results obtained using two different kernel functions and three different H$_0$ values.

To reconstruct the dark energy scalar field potential $V(\phi)$, we must choose priors and datasets. This paper aims to investigate how different priors and datasets affect the reconstruction results. We compare the reconstructed $V(\phi)$ using two priors: Planck 18 (P18) and WMAP nine-year (WMAP9y). We also compare the reconstructed $V(\phi)$ using three datasets: CC, BAO, and OHD (a combination of CC and BAO). Through this comparison, we can determine the impact of using different individual datasets (CC and BAO) on the reconstructed $V(\phi)$. Additionally, we can compare the differences between the individual datasets (CC or BAO) and the combined dataset (OHD) to understand the effect of combining different datasets on the reconstructed $V(\phi)$. With the advent of advancing technology, more $H(z)$ data points are being released, which can potentially optimize $V(\phi)$. We compare two models, Power Law and Free Field, to the reconstructed $V(\phi)$ by computing the reduced chi-square. This paper will also explore the impact of additional data points on the reconstruction results. The following sections will discuss these topics.

This work is structured as follows: In Sec. \ref{sec:Theory}, we derive the function of the dark energy scalar field potential $V(\phi)$. Sec. \ref{sec:Data and priors} compiles the CC dataset, BAO dataset, and priors in Table \ref{tab:table1}, \ref{tab:table2}, and \ref{tab:table3}, respectively. In Sec. \ref{sec:Reconstruction of the dark energy potential}, we reconstruct $V(\phi)$ using the CC, BAO, and OHD datasets via Gaussian Process. Two models, Power Law and Free Field, are compared to the reconstructed $V(\phi)$ by computing the reduced chi-square. We also simulate $H(z)$ to investigate how the accuracy of $V(\phi)$ reconstruction improves as we increase the number of $H(z)$ data points. Finally, Sec. \ref{sec:Discussions and conclusions} provides the conclusion and discussion of this paper.

\section{Theory} \label{sec:Theory}

Quintessence is a model of dark energy, characterized by a time-dependent scalar field. The properties of quintessence are represented by the dark energy scalar field potential $V(\phi)$. In order to reconstruct $V(\phi)$ with $H(z)$ datasets, it is necessary to derive the equation for $V(\phi)$. This can be done in two steps: first, by deriving the energy density $\rho_{\phi}$ and pressure $P_{\phi}$ of the scalar field; and second, by deriving $V(\phi)$.

\subsection{Energy density $\rho_{\phi}$ and pressure $P_{\phi}$}
\label{sec:Energy density and pressure}

The energy-momentum tensor for the scalar field $\phi(\vec{x},t)$ is
\begin{equation}
T^{\alpha}_{\beta}=g^{\alpha\nu}\frac{\partial\phi}{\partial x^{\nu}}\frac{\partial\phi}{\partial x^{\beta}}-g^{\alpha}_{\beta}\Bigl[\frac{1}{2}g^{\mu\nu}\frac{\partial\phi}{\partial x^{\mu}}\frac{\partial\phi}{\partial x^{\nu}}+V(\phi)\Bigr],
\label{eq:1}
\end{equation}
where $V(\phi)$ represents the dark energy scalar field potential. In this paper, we assume that the field is mostly homogeneous, and thus the energy-momentum for the zero-order part of it, $\phi^{(0)}(t)$, is
\begin{equation}
T^{\alpha(0)}_{\beta}=-g^{\alpha}_{0}g^{0}_{\beta}\Bigl(\frac{\mathrm{d}\phi^{(0)}}{\mathrm{d}t}\Bigr)^2+g^{\alpha}_{\beta}\Bigl[\frac{1}{2}\Bigl(\frac{\mathrm{d}\phi^{(0)}}{\mathrm{d}t}\Bigr)^2-V(\phi^{(0)})\Bigr].
\label{eq:2}
\end{equation}

Using Eq. (\ref{eq:2}), we can obtain the energy density $\rho_{\phi}$ and the pressure $P_{\phi}$. The energy density is given by the time-time component, $-T^{0}_0$. For the homogeneous field, the pressure $P_{\phi}$ is equal to $T^{i(0)}_i$, so
\begin{equation}
\rho_{\phi}=\frac{1}{2}\Bigl(\frac{\mathrm{d}\phi^{(0)}}{\mathrm{d}t}\Bigr)^2+V(\phi^{(0)}),
\label{eq:3}
\end{equation}
\begin{equation}
P_{\phi}=\frac{1}{2}\Bigl(\frac{\mathrm{d}\phi^{(0)}}{\mathrm{d}t}\Bigr)^2-V(\phi^{(0)}).
\label{eq:4}
\end{equation}

\subsection{dark energy scalar field potential $V(z)$}
\label{sec:dark energy scalar field potential}

Within the framework of general relativity, the evolution of the universe is described by the Friedmann equation and the acceleration equation. In our derivation, we use natural units. At late times in the universe, the radiation density is much smaller than other types of densities. Therefore, we can neglect the radiation density in the Friedmann equation, which is given by
\begin{equation}
H^2=\frac{8\pi G}{3}(\rho_{mat}+\rho_{\phi})-\frac{k}{a^2},
\label{eq:5}
\end{equation}
where $\rho_{mat}$ is the matter density in the universe, $k$ represents the curvature of the universe, and $a$ is the scalar factor. The acceleration equation is
\begin{equation}
\frac{\ddot{a}}{a}=-\frac{4\pi G}{3}(\rho_{mat}+\rho_{\phi}+3P_{\phi}).
\label{eq:6}
\end{equation}
By substituting $\rho_{\phi}$ and $P_{\phi}$ from Eq. (\ref{eq:3}) and (\ref{eq:4}) into Eq. (\ref{eq:5}) and (\ref{eq:6}), we obtain
\begin{equation}
H^2=\frac{8\pi G}{3}[\rho_{mat}+\frac{1}{2}\dot{\phi}^2+V(\phi)]-\frac{k}{a^2},
\label{eq:7}
\end{equation}
\begin{equation}
\frac{\ddot{a}}{a}=-\frac{4\pi G}{3}[\rho_{mat}+2\dot{\phi}^2-2V(\phi)].
\label{eq:8}
\end{equation}
To obtain the dark energy scalar field potential $V(z)$, we need to eliminate the $\dot{\phi}^2$ term in the equation. Note that $\frac{\ddot{a}}{a}=H^2+\dot{H}$. Multiplying Eq. (\ref{eq:7}) by 2 and adding it to Eq. (\ref{eq:8}) gives us
\begin{equation}
V(\phi)=\frac{1}{8\pi G}[3H^2+\dot{H}+\frac{2k}{a^2}]-\frac{\rho_{mat}}{2}.
\label{eq:9}
\end{equation}
In order to reconstruct the scalar field potential of dark energy in the redshift space, we need to convert $V(\phi)$ to $V(z)$. As we have assumed a homogeneous universe in this paper, the scalar field $\phi$ only varies with time t. The relationship between time and redshift is given by 
\begin{equation}
1+z=\frac{a(t_0)}{a(t_1)}.
\label{eq:add1}
\end{equation}
where z represents the redshift and $a(t_0)$ and $a(t_1)$ represent the scale factor at the present time and some past time, respectively. According to Eq. (\ref{eq:add1}), time can be expressed as a function of redshift, i.e. t=t(z), we can invert $V(\phi(t))$ to $V(\phi(z))$ and then to $V(z)$.
Eq. (\ref{eq:9}) can be rewritten as
\begin{eqnarray}
    V(z)=&&\frac{3H^2}{8\pi G}-\frac{1}{8\pi G}H(1+z)\frac{\mathrm{d}H}{\mathrm{d}z}-\frac{2H_0^2}{8\pi G}\Omega_{k0}(1+z)^2\nonumber\\
    &&-\frac{1}{2}\frac{3H_0^2}{8\pi G}\Omega_{M0}(1+z)^3.
\label{eq:10}
\end{eqnarray}

According to Eq. (\ref{eq:10}), in order to determine the values of $V(z)$ and $\sigma_V(z)$, it is necessary to know the values of $H$, $\mathrm{d}H/\mathrm{d}z$, $H_0$, $\Omega_{k0}$, and $\Omega_{M0}$ along with their 1$\sigma$ uncertainties (68\% confidence regions). We use the $H(z)$ dataset via the Gaussian Process to obtain $H$ and $\mathrm{d}H/\mathrm{d}z$ in the range of $z\in$ [0, 2.5], while H$_0$, $\Omega_{k0}$, and $\Omega_{M0}$ are chosen as priors. This paper uses the 1$\sigma$ uncertainty of $V(z)$, which is obtained by propagating the uncertainties. For a function $f=f(x_1,x_2,...,x_n)$, the variance formula of $f$ is given by
\begin{equation}
\sigma_f^2=\sum_{i}^{n}\Big(\frac{\partial f}{\partial x_i}\Big)^2{\sigma_{i}}^2+\sum_{i}^{n}\sum_{j(j\ne i)}^{n}\frac{\partial f}{\partial x_i}\frac{\partial f}{\partial x_j}\sigma_{ij}.
\label{eq:11}
\end{equation}
Where $\sigma_i$ represents the standard deviation of $x_i$, while $\sigma_{ij}$ denotes the covariance between $x_i$ and $x_j$. It is defined as:
\begin{equation}
\sigma_{ij}=\frac{1}{N}\sum_{k=1}^{N}(x_{ik}-\bar{x_i})(y_{ik}-\bar{y_i}).
\label{eq:sigmaij}
\end{equation}
$x_{ik}$ represents the kth value of $x_i$, and the total number of $x_i$ is denoted as N.

According to Eq. (\ref{eq:11}), the error equation for $V(z)$ is
\begin{equation}
    \sigma_V^2=\Big(\frac{\partial V}{\partial H}\Big)^2\sigma_{H}^2+\Big(\frac{\partial V}{\partial H'}\Big)^2\sigma_{H'}^2+\Big(\frac{\partial V}{\partial H_0}\Big)^2\sigma_{H_0}^2+\Big(\frac{\partial V}{\partial \Omega_{k0}}\Big)^2\sigma_{\Omega_{k0}}^2+\Big(\frac{\partial V}{\partial \Omega_{M0}}\Big)^2\sigma_{\Omega_{M0}}^2+2\frac{\partial V}{\partial H}\frac{\partial V}{\partial H'}\sigma_{HH'}.
\label{eq:12}
\end{equation}
Here, $H'$ represents $\mathrm{d}H/\mathrm{d}z$. $\sigma_H$, $\sigma_{H0}$, $\sigma_{H'}$, $\sigma_{\Omega_{k0}}$, and $\sigma_{\Omega_{M0}}$ represent the 1$\sigma$ uncertainty of $H$, H$_0$, $H'$, $\Omega_{k0}$, and $\Omega_{M0}$, respectively. $\sigma_{HH'}$ represents the covariance between $H$ and $H'$.

\section{Data and priors}
\label{sec:Data and priors}

The $H(z)$ dataset in Eq. (\ref{eq:10}) can be obtained from OHD. OHD comprises measurements of the Hubble parameter $H(z)$ at different redshifts, using various cosmological probes such as CC and BAO. In this paper, the CC and BAO datasets are used to reconstruct $H(z)$ via Gaussian process. CC provides a model-independent way to obtain $H(z)$ data directly calculated from the differential ages of galaxies
\begin{equation}
H(z)=-\frac{1}{1+z}\frac{\mathrm{d}z}{\mathrm{d}t}.
\label{eq:13}
\end{equation}
BAO can be measured from the correlation function \citep{2017MNRAS.469.3762W} or power spectrum \citep{2012MNRAS.427.3435A} of the galaxy distribution. The Hubble parameter $H(z)$, detectable in the direction of the BAO's line-of-sight, can be parameterized as
\begin{equation}
\alpha_{\parallel}=\frac{H^{fid}(z)r^{fid}_d}{H(z)r_d}.
\label{eq:14}
\end{equation}
The CC and BAO data used in this paper are compiled in Table \ref{tab:table1} and \ref{tab:table2}, respectively. We reconstruct $V(z)$ in Eq. (\ref{eq:11}) using the CC, BAO, and OHD (CC+BAO) datasets. We compare the influence of different individual datasets (CC and BAO) on the reconstructed $V(z)$. Additionally, we compare the impact of using individual datasets (CC or BAO) versus the combined dataset (OHD) on the reconstructed $V(z)$.

The OHD data points are collected from different galaxies using various methods. We assume that data points obtained through different methods are independent, with no covariance between them. Within the OHD dataset, there are 15 data points proposed by \cite{2012JCAP...08..006M}, \cite{ 2015MNRAS.450L..16M}, and \cite{ 2016JCAP...05..014M}, all obtained through the same method. The researchers responsible for these data points have introduced a method known as the \href{https://gitlab.com/mmoresco/CCcovariance}{Cosmic chronometers covariance} \citep{2020ApJ...898...82M} to estimate the covariance between them. We take this covariance matrix into account when reconstructing the dark energy scalar field potential in Section \ref{sec:Reconstruction}.

%------------------table1------------
\begin{table}
\begin{minipage}{\textwidth}
	\centering\caption{Compiled CC data}
	
	\begin{tabular*}{.52\textwidth}{lccc}
		\hline\hline
		~~~Redshift $z$ &\hspace*{0em} $H(z)\footnote{$H(z)$ figures are in the unit of kms$^{-1}$ Mpc$^{-1}$}$$\pm 1\sigma$ error &\hspace*{0em}References \\
		
		\hline
        ~~~~~~0.07&\hspace*{0em}69$\pm$19.6 & \hspace*{0em}\cite{2014RAA....14.1221Z}\\
        ~~~~~~0.1&\hspace*{0em}69$\pm$12 & \hspace*{0em}\cite{2010JCAP...02..008S}\\
        ~~~~~~0.12&\hspace*{0em}68.6$\pm$26.2 & \hspace*{0em}\cite{2014RAA....14.1221Z}\\
        ~~~~~~0.17&\hspace*{0em}83$\pm$8 & \hspace*{0em}\cite{2010JCAP...02..008S}\\
        ~~~~~~0.1791&\hspace*{0em}75$\pm$4 & \hspace*{0em}\cite{2012JCAP...08..006M}\\
        ~~~~~~0.1993&\hspace*{0em}75$\pm$5 & \hspace*{0em}\cite{2012JCAP...08..006M}\\
        ~~~~~~0.2&\hspace*{0em}72.9$\pm$29.6 & \hspace*{0em}\cite{2014RAA....14.1221Z}\\
        ~~~~~~0.27&\hspace*{0em}77$\pm$14 & \hspace*{0em}\cite{2010JCAP...02..008S}\\
        ~~~~~~0.28&\hspace*{0em}88.8$\pm$36.6 & \hspace*{0em}\cite{2014RAA....14.1221Z}\\
        ~~~~~~0.3519&\hspace*{0em}83$\pm$14 & \hspace*{0em}\cite{2012JCAP...08..006M}\\
        ~~~~~~0.382&\hspace*{0em}83$\pm$13.5 & \hspace*{0em}\cite{2016JCAP...05..014M}\\
        ~~~~~~0.4&\hspace*{0em}95$\pm$17 & \hspace*{0em}\cite{2010JCAP...02..008S}\\
        ~~~~~~0.4004&\hspace*{0em}77$\pm$10.2 & \hspace*{0em}\cite{2016JCAP...05..014M}\\
        ~~~~~~0.4247&\hspace*{0em}87.1$\pm$11.2 & \hspace*{0em}\cite{2016JCAP...05..014M}\\
        ~~~~~~0.4497&\hspace*{0em}92.8$\pm$12.9 & \hspace*{0em}\cite{2016JCAP...05..014M}\\
        ~~~~~~0.47&\hspace*{0em}89$\pm$49.6 & \hspace*{0em}\cite{2017MNRAS.467.3239R}\\
        ~~~~~~0.4783&\hspace*{0em}80.9$\pm$9 & \hspace*{0em}\cite{2016JCAP...05..014M}\\
        ~~~~~~0.48&\hspace*{0em}97$\pm$62 & \hspace*{0em}\cite{2010JCAP...02..008S}\\
        ~~~~~~0.5929&\hspace*{0em}104$\pm$13 & \hspace*{0em}\cite{2012JCAP...08..006M}\\
        ~~~~~~0.6797&\hspace*{0em}92$\pm$8 & \hspace*{0em}\cite{2012JCAP...08..006M}\\
        ~~~~~~0.7812&\hspace*{0em}105$\pm$12 & \hspace*{0em}\cite{2012JCAP...08..006M}\\
        ~~~~~~0.8&\hspace*{0em}113.1$\pm$25.22 & \hspace*{0em}\cite{2023ApJS..265...48J}\\
        ~~~~~~0.8754&\hspace*{0em}125$\pm$17 & \hspace*{0em}\cite{2012JCAP...08..006M}\\
        ~~~~~~0.88&\hspace*{0em}90$\pm$40 & \hspace*{0em}\cite{2010JCAP...02..008S}\\
        ~~~~~~0.9&\hspace*{0em}117$\pm$23 & \hspace*{0em}\cite{2010JCAP...02..008S}\\
        ~~~~~~1.037&\hspace*{0em}154$\pm$20 & \hspace*{0em}\cite{2012JCAP...08..006M}\\
        ~~~~~~1.3&\hspace*{0em}168$\pm$17 & \hspace*{0em}\cite{2010JCAP...02..008S}\\
        ~~~~~~1.363&\hspace*{0em}160$\pm$33.6 & \hspace*{0em}\cite{2015MNRAS.450L..16M}\\
        ~~~~~~1.43&\hspace*{0em}177$\pm$18 & \hspace*{0em}\cite{2010JCAP...02..008S}\\
        ~~~~~~1.53&\hspace*{0em}140$\pm$14 & \hspace*{0em}\cite{2010JCAP...02..008S}\\
        ~~~~~~1.75&\hspace*{0em}202$\pm$40 & \hspace*{0em}\cite{2010JCAP...02..008S}\\
        ~~~~~~1.965&\hspace*{0em}186.5$\pm$50.4 & \hspace*{0em}\cite{2015MNRAS.450L..16M}\\
		\hline\hline
	\end{tabular*}
 	\label{tab:table1}
\end{minipage}
\end{table}

%------------------table1------------

%------------------table2------------
\begin{table}
\begin{minipage}{\textwidth}
	\centering\caption{Compiled BAO data}
	
	\begin{tabular*}{.52\textwidth}{lccc}
		\hline\hline
		~~~Redshift $z$ &\hspace*{0em} $H(z)\pm 1\sigma$ error &\hspace*{0em}References \\
		
		\hline
        ~~~~~~0.24&\hspace*{0em}82.37$\pm$3.94& \hspace*{0em}\cite{2009MNRAS.399.1663G} \\
        ~~~~~~0.3&\hspace*{0em}78.83$\pm$6.58& \hspace*{0em}\cite{2014MNRAS.439.2515O} \\
        ~~~~~~0.31&\hspace*{0em}78.39$\pm$5.46& \hspace*{0em}\cite{2017MNRAS.469.3762W} \\
        ~~~~~~0.35&\hspace*{0em}88.1$\pm$9.45& \hspace*{0em}\cite{2013MNRAS.435..255C} \\
        ~~~~~~0.36&\hspace*{0em}80.16$\pm$4.37& \hspace*{0em}\cite{2017MNRAS.469.3762W} \\
        ~~~~~~0.38&\hspace*{0em}81.74$\pm$3.4& \hspace*{0em}\cite{2017MNRAS.470.2617A} \\
        ~~~~~~0.43&\hspace*{0em}89.36$\pm$4.89& \hspace*{0em}\cite{2009MNRAS.399.1663G} \\
        ~~~~~~0.44&\hspace*{0em}85.48$\pm$8.59& \hspace*{0em}\cite{2012MNRAS.425..405B} \\
        ~~~~~~0.51&\hspace*{0em}90.67$\pm$3.66& \hspace*{0em}\cite{2017MNRAS.470.2617A} \\
        ~~~~~~0.52&\hspace*{0em}94.61$\pm$4.2& \hspace*{0em}\cite{2017MNRAS.469.3762W} \\
        ~~~~~~0.56&\hspace*{0em}93.59$\pm$3.96& \hspace*{0em}\cite{2017MNRAS.469.3762W} \\
        ~~~~~~0.57&\hspace*{0em}96.59$\pm$8.76& \hspace*{0em}\cite{2014MNRAS.439...83A} \\
        ~~~~~~0.59&\hspace*{0em}98.75$\pm$4.66& \hspace*{0em}\cite{2017MNRAS.469.3762W} \\
        ~~~~~~0.6&\hspace*{0em}90.96$\pm$7.04& \hspace*{0em}\cite{2012MNRAS.425..405B} \\
        ~~~~~~0.61&\hspace*{0em}97.59$\pm$3.97& \hspace*{0em}\cite{2017MNRAS.470.2617A} \\
        ~~~~~~0.64&\hspace*{0em}99.09$\pm$4.53& \hspace*{0em}\cite{2017MNRAS.469.3762W} \\
        ~~~~~~0.73&\hspace*{0em}100.69$\pm$8.03& \hspace*{0em}\cite{2012MNRAS.425..405B} \\
        ~~~~~~2.33&\hspace*{0em}223.99$\pm$11.12& \hspace*{0em}\cite{2017AA...603A..12B} \\
        ~~~~~~2.34&\hspace*{0em}222.105$\pm$10.38& \hspace*{0em}\cite{2015AA...574A..59D} \\
        ~~~~~~2.36&\hspace*{0em}226.24$\pm$11.18& \hspace*{0em}\cite{2014JCAP...05..027F} \\
		\hline\hline
	\end{tabular*}
	\label{tab:table2} 
\end{minipage}
\end{table}

%------------------table2------------

According to Eq. (\ref{eq:10}) and (\ref{eq:12}), it is necessary to have priors on H$_0$, $\Omega_{k0}$, and $\Omega_{M0}$ in order to reconstruct the Hubble parameter $H(z)$. To compare the effects of different priors on the results, we choose two different priors, Planck 18 (P18) and WMAP nine-year (WMAP9y), to reconstruct $H(z)$. The priors are compiled in Table \ref{tab:table3}. We choose two priors and three datasets for the reconstruction of $V(z)$, resulting in six different combinations of dataset$\_$prior. These reconstructed results are shown in Fig. \ref{fig:CC_p18}, \ref{fig:cc_WMAP9y}, \ref{fig:bao_P18}, \ref{fig:bao_WMAP9y}, \ref{fig:OHD_p18_differentcolor}, \ref{fig:ohd_WMAP9y_differentcolor}.

%------------------table3------------
\begin{table}
\begin{minipage}{\textwidth}
	\centering\caption{Compiled priors}
	
	\begin{tabular*}{.75\textwidth}{ccccc}
		\hline\hline
		~~~Prior &\hspace*{0em} $H_0$ &\hspace*{0em}$\Omega_{k0}$ &\hspace*{0em}$\Omega_{M0}$ &\hspace*{0em}References \\
		
		\hline
            P18 & 67.4±0.5 & 0.001±0.002 & 0.315±0.007 & \cite{2020AA...641A...6P}\\
            WMAP9y & 70.0±2.2 & -0.037±0.043 & 0.279±0.025 & \cite{2013ApJS..208...20B}\\
		\hline\hline
	\end{tabular*}
	\label{tab:table3} 
\end{minipage}
\end{table}

%------------------table3------------

\section{Reconstruction of the dark energy scalar field potential}
\label{sec:Reconstruction of the dark energy potential}

\subsection{Reconstruction}
\label{sec:Reconstruction}

Gaussian Process is a popular method for reconstructing a function without assuming any parameters in the function \citep{2016PhRvD..93d3517C,2021ApJ...915..123S,2014PhRvD..89b3503Y}. Gaussian Process can describe the observational data by a distribution function $f(z)$ with mean value and error at each point $z$. The reconstructed functions at different points $z$ and $\tilde{z}$ are correlated by a covariance function $k(z, \tilde{z})$. There are many covariance functions available \citep{2006gpml.book.....R,2013arXiv1311.6678S}. In this paper, we use the Squared Exponential covariance function, which is widely used in the reconstruction of $H(z)$ by Gaussian Process  \citep{2012JCAP...06..036S, 2014MNRAS.441L..11B, 2021ApJ...915..123S}. The Squared Exponential covariance function is given by
\begin{equation}
    k(z, \tilde{z}) = \sigma^2_f \mathrm{exp} \left( - \frac{(z - \tilde{z})^2}{2l^2} \right).
    \label{eq:SquaredExponentia}
\end{equation}  
This function depends on two hyperparameters: $\sigma_f$ and $l$. These hyperparameters determine the amplitude and the coherence length of the function, respectively. In this paper, the Gaussian Process is used to reconstruct the $H(z)$ function from the data points listed in Table \ref{tab:table1} and Table \ref{tab:table2}. Compared to other reconstruction methods like Artificial Neural Networks (ANN), the distribution function is a Gaussian distribution, allowing it to be differential at any point $z$. This property can be utilized to easily reconstruct the first derivative of the Hubble parameter $H(z)$ with respect to z, denoted as $H'(z)$. The code used in this paper is Gaussian Processes in Python (GAPP) proposed by \cite{2012JCAP...06..036S}.

\begin{figure*}
\includegraphics[width=17cm]{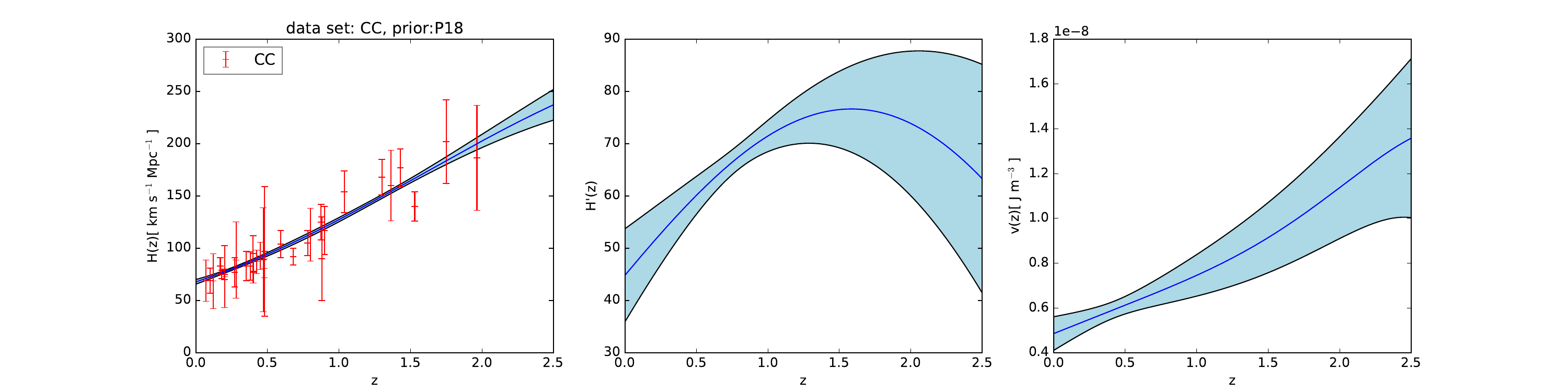}% Here is how to import EPS art
\caption{\label{fig:CC_p18} 
Gaussian Process reconstruction of $H(z)$ (left) from CC data set, $H'(z)$ (middle), and $V(z)$ (right) with P18 prior. The blue-shaded region is the 1$\sigma$ uncertainty of the reconstruction.} 
\end{figure*}

\begin{figure*}
\includegraphics[width=17cm]{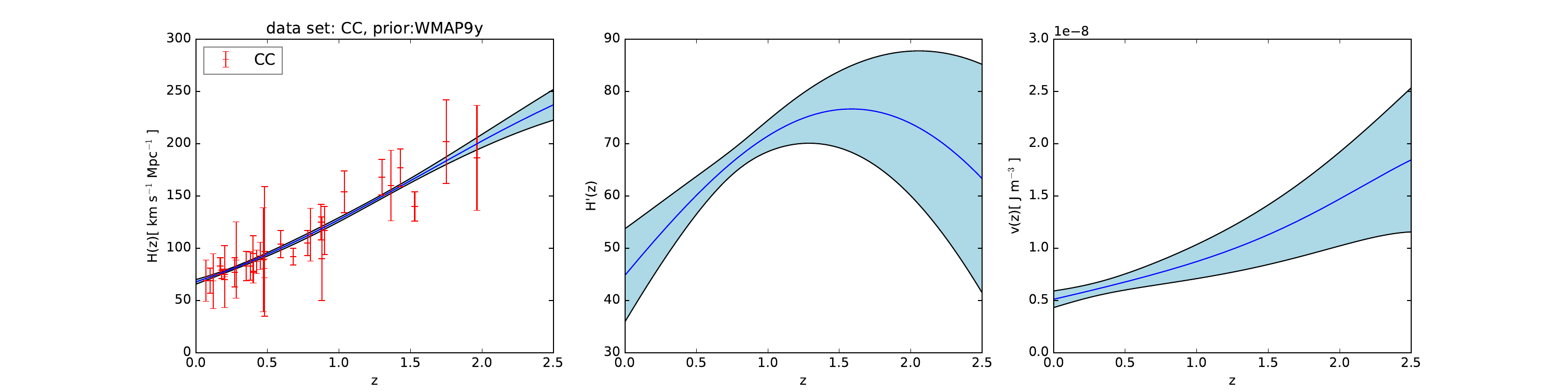}% Here is how to import EPS art
\caption{\label{fig:cc_WMAP9y} Gaussian Process reconstruction of $H(z)$ (left) from CC data set, $H'(z)$ (middle), and $V(z)$ (right) with WMAP9y prior. The blue-shaded region is the 1$\sigma$ uncertainty of the reconstruction.}
\end{figure*}

\begin{figure*}
\includegraphics[width=17cm]{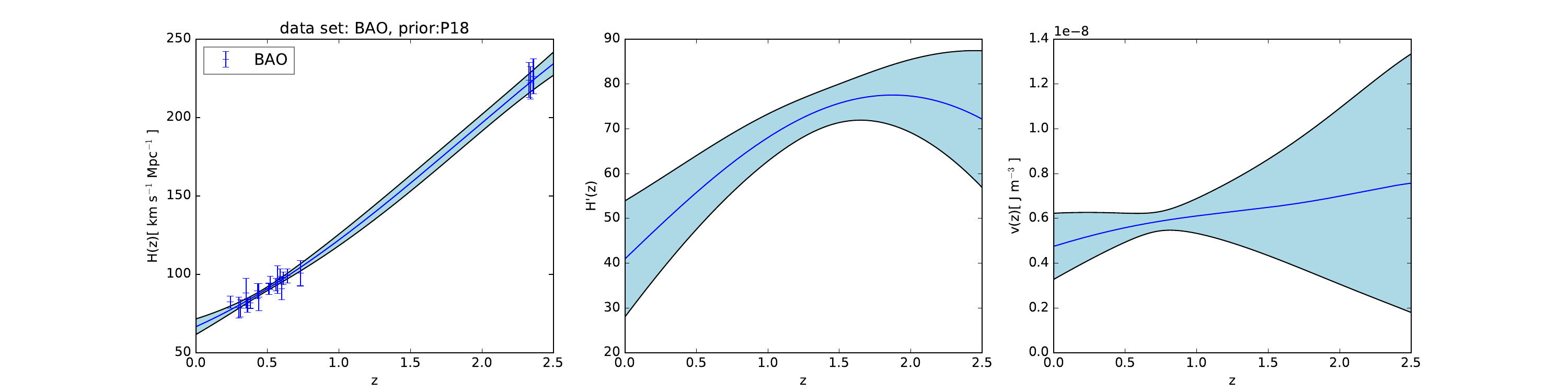}% Here is how to import EPS art
\caption{\label{fig:bao_P18} Gaussian Process reconstruction of $H(z)$ (left) from BAO data set, $H'(z)$ (middle), and $V(z)$ (right) with P18 prior. The blue-shaded region is the 1$\sigma$ uncertainty of the reconstruction.}
\end{figure*}

\begin{figure*}
\includegraphics[width=17cm]{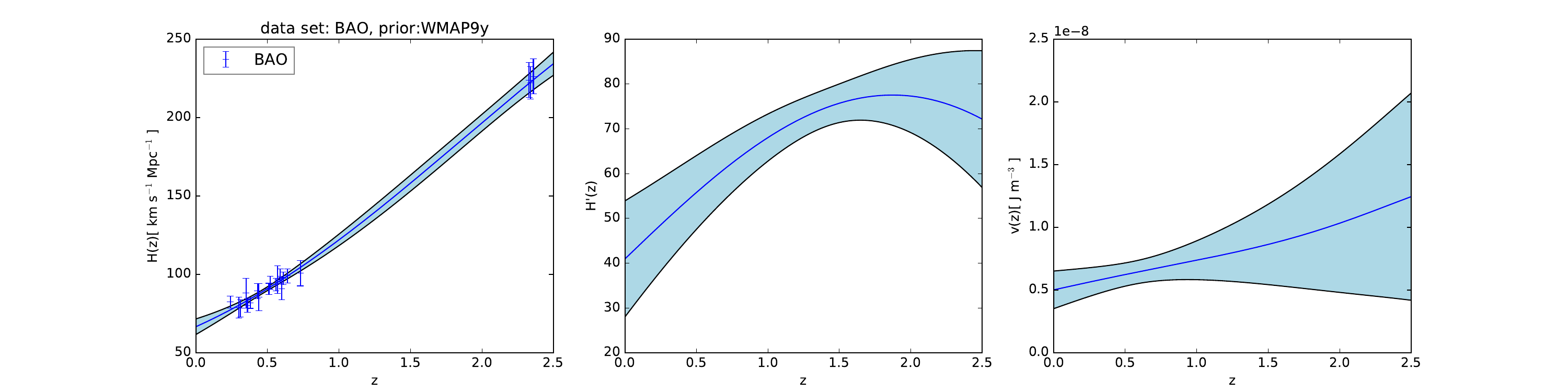}% Here is how to import EPS art
\caption{\label{fig:bao_WMAP9y} Gaussian Process reconstruction of $H(z)$ (left) from BAO data set, $H'(z)$ (middle), and $V(z)$ (right) with WMAP9y prior. The blue-shaded region is the 1$\sigma$ uncertainty of the reconstruction.}
\end{figure*}

\begin{figure*}
\includegraphics[width=17cm]{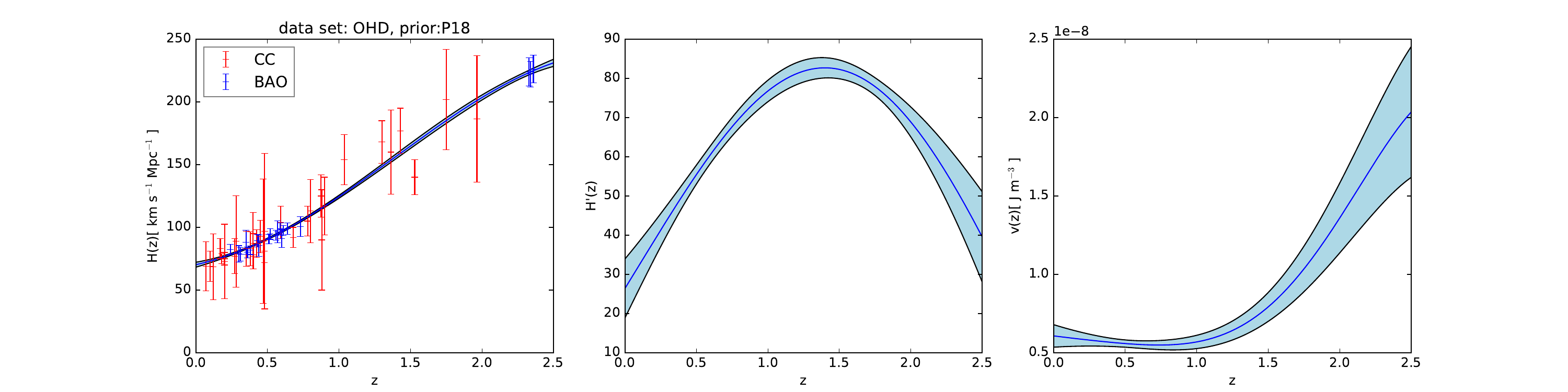}% Here is how to import EPS art
\caption{\label{fig:OHD_p18_differentcolor} Gaussian Process reconstruction of $H(z)$ (left) from OHD data set, $H'(z)$ (middle), and $V(z)$ (right) with P18 prior. The OHD data set contains CC data points in red and BAO data points in blue. The blue-shaded region is the 1$\sigma$ uncertainty of the reconstruction.}
\end{figure*}

\begin{figure*}
\includegraphics[width=17cm]{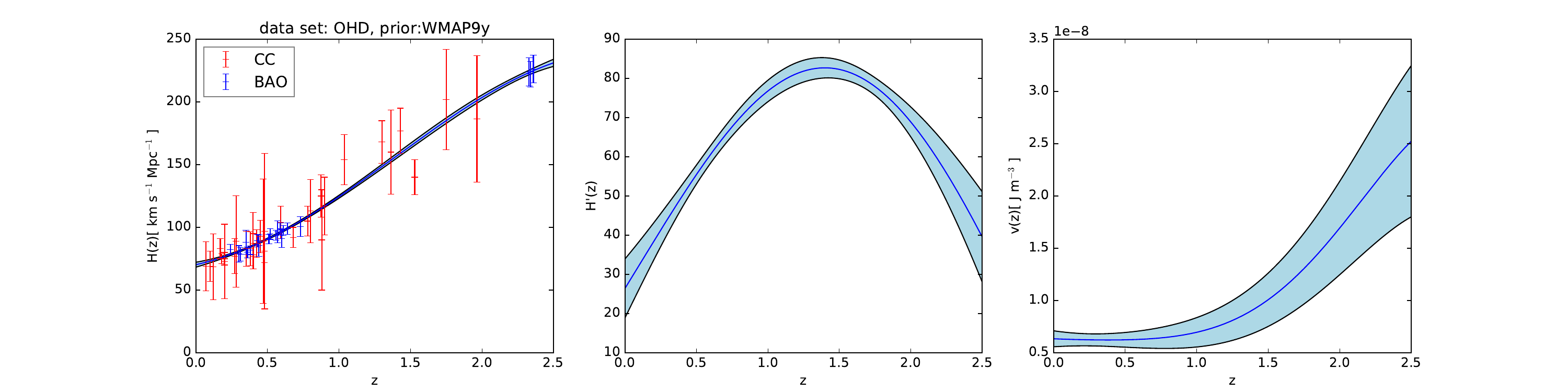}% Here is how to import EPS art
\caption{\label{fig:ohd_WMAP9y_differentcolor} Gaussian Process reconstruction of $H(z)$ (left) from OHD data set, $H'(z)$ (middle), and $V(z)$ (right) with WMAP9y prior. The OHD data set contains CC data points in red and BAO data points in blue. The blue-shaded region is the 1$\sigma$ uncertainty of the reconstruction.}
\end{figure*}

To reconstruct the dark energy scalar field potential $V(z)$, we need the reconstructed $H(z)$ and $H'(z)$ values with 1$\sigma$ uncertainty. Additionally, we require the a priors of H$_0$, $\Omega_{k0}$, and $\Omega_{M0}$ as listed in Table \ref{tab:table3}, along with the covariance $\sigma_{HH'}$ between $H(z)$ and $H'(z)$ in Eq. (\ref{eq:12}). The use of the Gaussian Process enables us to reconstruct $H(z)$ and $H'(z)$ from the observational data. To determine the impact of different datasets and priors on $V(z)$, we consider three datasets (CC, BAO, and OHD), and two different priors (P18 and WMAP9y). This results in six different reconstructed $V(z)$ outcomes: CC+P18 as shown in Fig. \ref{fig:CC_p18}, CC+WMAP9y as shown in Fig. \ref{fig:cc_WMAP9y}, BAO+P18 as shown in Fig. \ref{fig:bao_P18}, BAO+WMAP9y as shown in Fig. \ref{fig:bao_WMAP9y}, OHD+P18 as shown in Fig. \ref{fig:OHD_p18_differentcolor}, and OHD+WMAP9y as shown in Fig. \ref{fig:ohd_WMAP9y_differentcolor}.

%%%%%%%%%%%%%%%%%%%%%%%response to JCAP%%%%%%%%%%%%%%%%

\begin{figure}
\centering
\includegraphics[width=19cm]{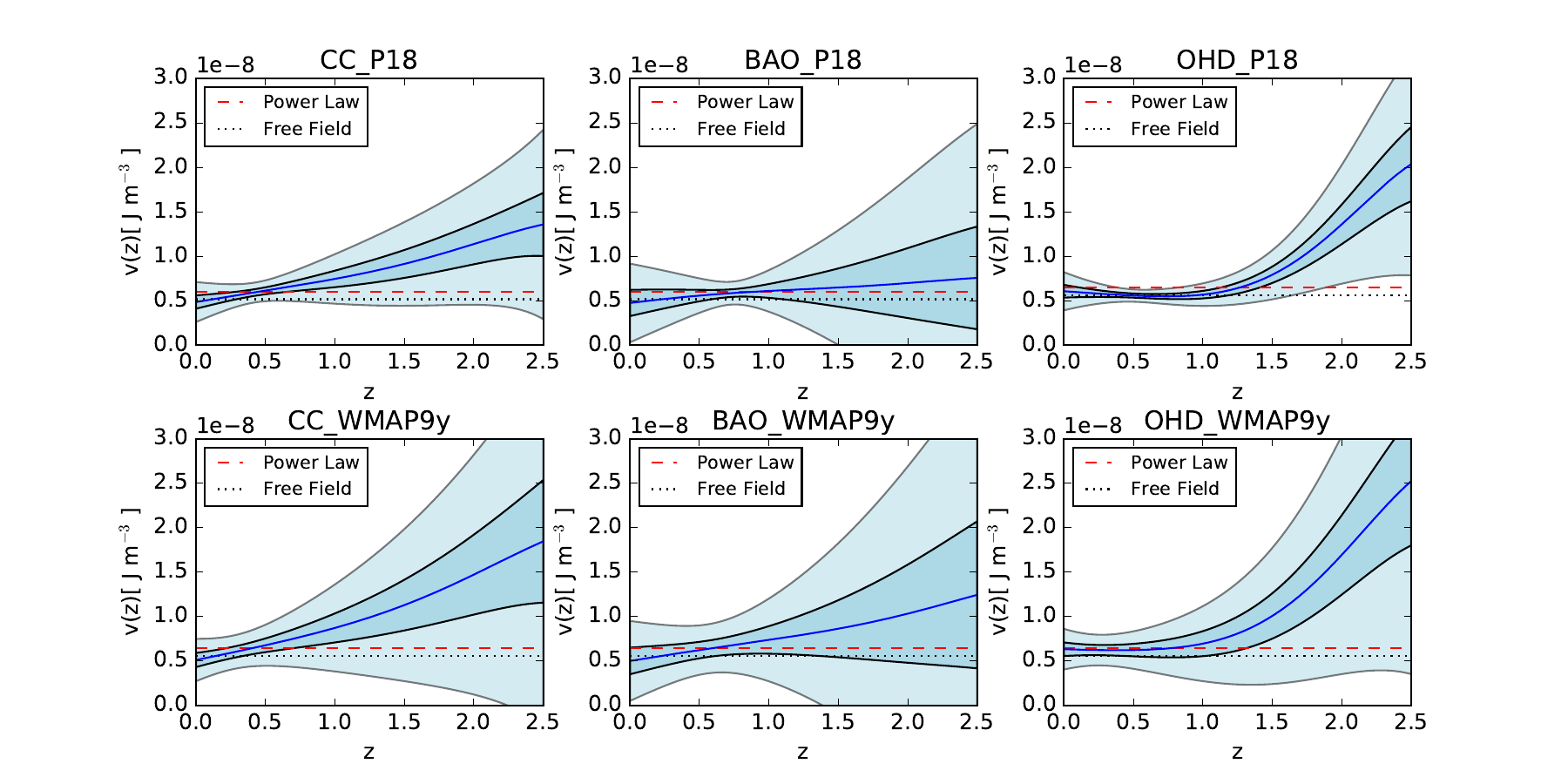}% Here is how to import EPS art
\caption{\label{fig:theoritical_recnstruction} %same dataset; different prior
Gaussian process reconstruction of $V(z)$ from six different datasets and priors: CC with prior P18 (upper left), BAO with prior P18 (upper center), OHD with prior P18 (upper right), CC with prior WMAP9y (bottom left), BAO with prior WMAP9y (bottom center), and OHD with prior WMAP9y (bottom right). The shaded blue regions represent the $1\sigma$ and $3\sigma$ uncertainty of the reconstruction. The red dashed line and the black dotted line represent the dark energy scalar field models Power Law and Free Field, respectively.}
\end{figure}

The reconstructed $V(z)$ can validate and compare various models of the dark energy scalar field. In this study, we have selected two specific models \citep{2022JCAP...11..037J}: the Free Field (FF) model \citep{1991PhRvD..44..352R, 2009IJMPD..18..621U} and the Power Law (PL) \citep{1988ApJ...325L..17P, 1988PhRvD..37.3406R} model:
\begin{equation}
V(\phi)_{FF}= \frac{3\mathrm{H}_0^2}{8\pi G}\frac{\mu^2}{2}\phi^2,
\label{eq:FreeField}
\end{equation}
where $\mu\equiv\frac{m}{H_0}$.
\begin{equation}
V(\phi)_{PL}= \frac{3\mathrm{H}_0^2}{8\pi G}\frac{K\phi^{-\alpha}}{2},
\label{eq:PowerLaw}
\end{equation}
\begin{equation}
K = \frac{8}{3}\frac{\alpha+4}{\alpha+2}[\frac{2}{3}\alpha(\alpha+2)]^{\alpha/2}.
\label{eq:K}
\end{equation}
The comparison between the reconstructed $V(z)$ and these models is illustrated in Fig. \ref{fig:theoritical_recnstruction}.

As depicted in Fig. \ref{fig:theoritical_recnstruction}, the Power Law and Free Field models fall within the $3\sigma$ range of reconstructed $V(z)$ for five dataset$\_$prior conditions (CC$\_$P18, CC$\_$WMAP9y, BAO$\_$P18, BAO$\_$WMAP9y, and OHD$\_$WMAP9y) in this paper. However, for the dataset: OHD, prior: P18 condition, at high redshift ($z\gtrsim 1.5$), these models fall outside the $3\sigma$ range of the reconstructed $V(z)$. This implies that the Power Law model and Free Field model are excluded at high redshift in the dataset: OHD, prior: P18 condition. Based on the figure, at low redshift ($z \lesssim 0.5$ for CC$\_$P18 and CC$\_$WMAP9y; $z \lesssim 1$ for OHD$\_$WMAP9y), both models exhibit a high level of compatibility, falling within a deviation of less than $1 \sigma$ from the reconstructed $V(z)$. For the remaining redshift range, compatibility is within $3 \sigma$. These models fit the reconstructed $V(z)$ better at low redshifts. Meanwhile, for BAO$\_$P18 and BAO$\_$WMAP9y conditions across the entire redshift range from 0 to 2.5, both models are almost within 1σ of the reconstructed $V(z)$, indicating a high level of compatibility with the reconstructed $V(z)$ under these conditions.

To assess which model provides a better fit to the reconstructed $V(z)$ across different datasets and priors depicted in Fig. \ref{fig:theoritical_recnstruction}, we computed the reduced chi-square ($\chi^2_{red}$) values between the reconstructed $V(z)$ and various models. The formula of $\chi^2_{red}$ is:
\begin{equation}
\chi^2_{red} = \frac{1}{N-p} \Sigma \frac{V_m-V_r}{\sigma_{V_r}^2},
\label{eq:chi2}
\end{equation}
where N represents the number of data points and p represents the number of parameters in the model. $V_m$ represents the $V(z)$ provided by the theoretical model. $V_r$ and $\sigma_{V_r}^2$ denote the reconstructed $V(z)$ and its 1$\sigma$ uncertainty, respectively. We computed the $\chi^2_{red}$ between the reconstructed $V(z)$ and the theoretical model's $V(z)$, where the z-values are the observed z-values as presented in Table \ref{tab:table1} and \ref{tab:table2}. Consequently, N equals 32 for CC$\_$P18 and CC$\_$WMAP9y, 20 for BAO$\_$P18 and BAO$\_$WMAP9y, and 52 for OHD$\_$WMAP9y. The result of $\chi^2_{red}$ is compiled in Table \ref{tab:label4}.

 In the first column of Table \ref{tab:label4}, dataset$\_$prior indicates the dataset and prior used for reconstructing $V(z)$. Table \ref{tab:label4} reveals that cross all dataset$\_$priors (CC$\_$P18, CC$\_$WMAP9y, BAO$\_$P18, BAO$\_$WMAP9y, OHD$\_$WMAP9y), Power Law model consistently has lower $\chi^2_{red}$ values compared to Free Field model. This suggests that, in general, the Power Law model provides a better fit to the reconstructed $V(z)$ data. For both Power Law model and Free Field model, the BAO$\_$P18 dataset$\_$prior consistently results in lower $\chi^2_{red}$ values compared to other dataset$\_$prior conditions. This indicates that the BAO$\_$P18 dataset$\_$prior is a better fit for both models when considering the $\chi^2_{red}$ values alone. For the CC and BAO datasets, the choice of prior (P18 vs. WMAP9y) significantly affects the fit quality, with P18 generally providing better results.

%------------------table4---------------

\begin{table}
    \centering\caption{$\chi^2_{red}$ Comparison: Reconstructed $V(z)$ vs. Different Models}
    \begin{tabular}{|c|c|c|}
    \hline
    dataset$\_$prior & model & $\chi^2_{red}$ \\
    \hline
    \multirow{2}{*}{\makecell{CC$\_$P18}} & Power Law &  $4.92$ \\
    \cline{2-3}
    & Free Field & $8.13$ \\
    \hline
    \multirow{2}{*}{\makecell{CC$\_$WMAP9y}} & Power Law & $12.18$ \\
    \cline{2-3}
    & Free Field & $17.59$ \\
    \hline
    \multirow{2}{*}{\makecell{BAO$\_$P18}} & Power Law &  $0.24$ \\
    \cline{2-3}
    & Free Field & $0.41$ \\
    \hline
    \multirow{2}{*}{\makecell{BAO$\_$WMAP9y}} & Power Law &  $2.26$ \\
    \cline{2-3}
    & Free Field & $3.33$ \\    
    \hline    
    \multirow{2}{*}{\makecell{OHD$\_$WMAP9y}} & Power Law & $34.97$ \\
    \cline{2-3}
    & Free Field & $41.40$ \\
    \hline
    \end{tabular}
    \label{tab:label4}
\end{table}

%------------------table4---------------

%%%%%%%%%%%%%%%%%%%%%%%response to JCAP%%%%%%%%%%%%%%%%%%%%%%

\subsection{Comparison}
\label{sec:Comparison}

\begin{figure}[H]
\centering
\includegraphics[width=19cm]{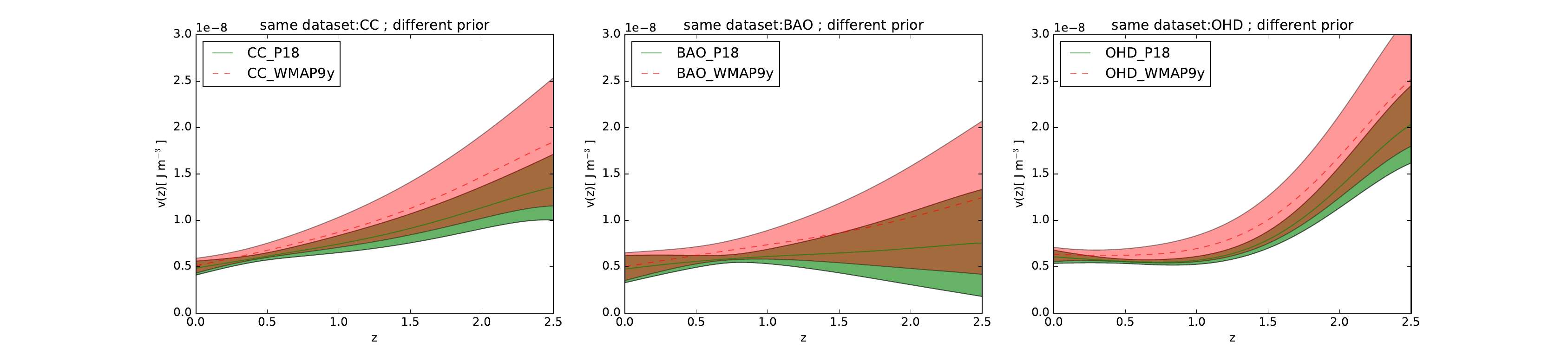}% Here is how to import EPS art
\caption{\label{fig:same dataset; different prior} %same dataset; different prior
Comparison of the reconstructed $V(z)$ from the CC (left), BAO (center), and OHD (right) datasets using different priors. The solid green line represents the mean of $V(z)$ with P18 prior, while the dashed red line represents the mean of $V(z)$ with WMAP9y prior. The green and red shaded regions correspond to the 1$\sigma$ uncertainty of $V(z)$ with P18 and WMAP9y, respectively.}
\end{figure}

\begin{figure}[H]
\centering
\includegraphics[width=13cm]{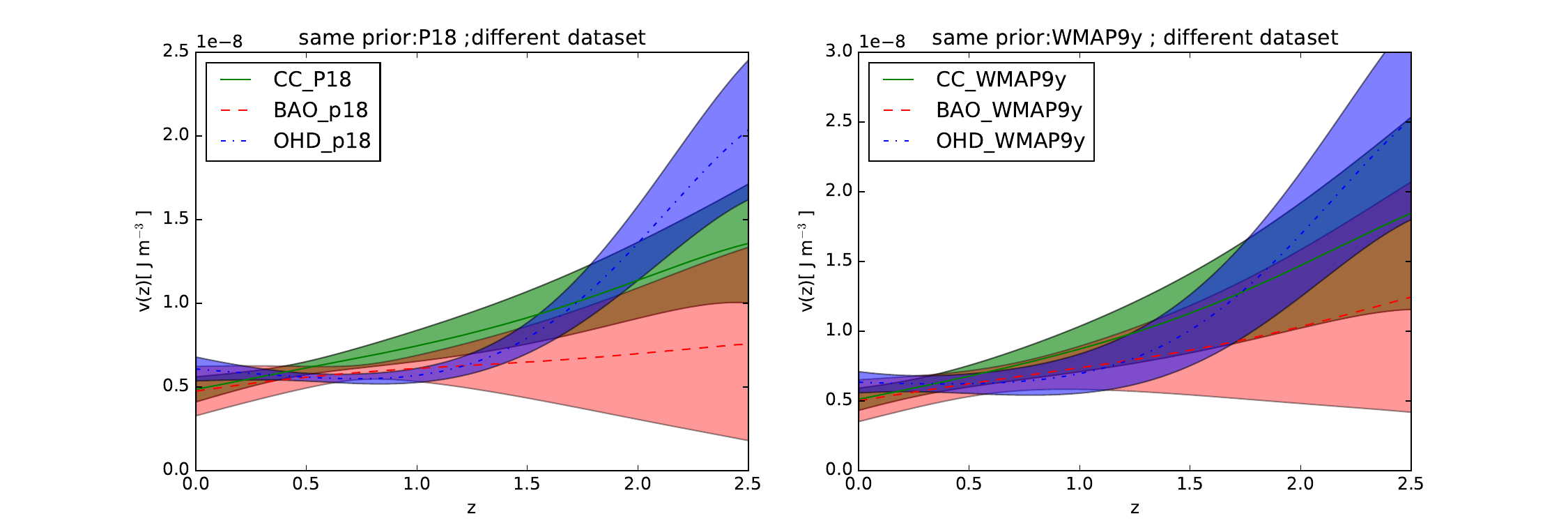}% Here is how to import EPS art
\caption{\label{fig:same prior; different dataset} %same prior; different dataset
Comparison of the reconstructed $V(z)$ using the same prior, P18 (left) and WMAP9y (right), from the different datasets. The solid green line represents the mean $V(z)$ reconstructed from the CC dataset, the dashed red line represents the mean $V(z)$ reconstructed from the BAO dataset, and the dash-dotted blue line represents the mean $V(z)$ reconstructed from the OHD dataset. The green, red, and blue shaded regions represent the 1$\sigma$ uncertainty of $V(z)$ from the CC, BAO, and OHD datasets, respectively.}
\end{figure}

In this paper, we aim to compare the effects of different priors and datasets on the reconstructed $V(z)$. We divide the six $V(z)$ results mentioned in Sec. \ref{sec:Reconstruction} into two categories. The first category is the comparison of the same dataset with different priors, as shown in Fig. \ref{fig:same dataset; different prior}. We observe that the choice of priors has a substantial effect on the reconstructed $V(z)$, and it is worth noting that the P18 prior consistently results in narrower uncertainty regions across all datasets, indicating more precise reconstructions compared to the WMAP9y prior. The second category is the comparison of the same prior with different datasets: CC, BAO, and OHD  (which combines the CC and BAO datasets), as shown in Fig. \ref{fig:same prior; different dataset}. We observe a significant difference in the reconstructed $V(z)$ based on different datasets. Different datasets lead to distinct trends in the reconstructed $V(z)$. The CC dataset consistently exhibits a narrow uncertainty band, indicating higher precision compared to the BAO dataset. CC results are consistent with BAO within $1.19 \sigma$ for P18 prior (left panel of Fig. \ref{fig:same prior; different dataset}) and $0.62 \sigma$ for WMAP9y prior (right panel of Fig. \ref{fig:same prior; different dataset}), justifying their combination to derive results for the OHD dataset. The combined OHD dataset shows a trend distinct from CC and BAO, highlighting the impact of combining datasets. These results indicate that obtaining more accurate $H(z)$ data points and precise priors will enhance the accuracy and reliability of the reconstructed $V(z)$.

\subsection{Simulation}
\label{sec:Simulation}

To quantify the influence of the reconstructed $V(z)$ results based on the number of $H(z)$ data points, we simulate additional $H(z)$ data points using the available $H(z)$ datasets in Table \ref{tab:table1} and \ref{tab:table2}. The simulation is based on the principle of generating $H_{sim}(z)$ data points that are similar to the available datasets. The simulation of $H_{sim}(z)$ consists of two main steps. In the first step, we simulate $H_{m}(z)$, which represents the mean value of the simulated Hubble parameter $H_{sim}(z)$ at different redshifts $z$. In the second step, we generate the uncertainty of the Hubble parameter, $\sigma_{H}(z)$. The $H_{sim}(z)$ at redshift $z_{sim}$ is
\begin{equation}
H_{sim}(z_{sim})=H_m(z_{sim})+\sigma_{H}(z_{sim}).
\label{eq:15}
\end{equation}

To quantify the improvement of the $V(z)$ result by increasing the number of $H(z)$ data points, we double the number of $H(z)$ data points, which required simulating 32 CC and 20 BAO data points similar to the datasets compiled in Table \ref{tab:table1} and \ref{tab:table2}. In this paper, we simulate redshift $z_{sim}$ with the same density distribution as the observational datasets. This approach is more reasonable than a uniform distribution, considering the current technology and available datasets.

\begin{figure}[H]
\includegraphics[width=15cm]{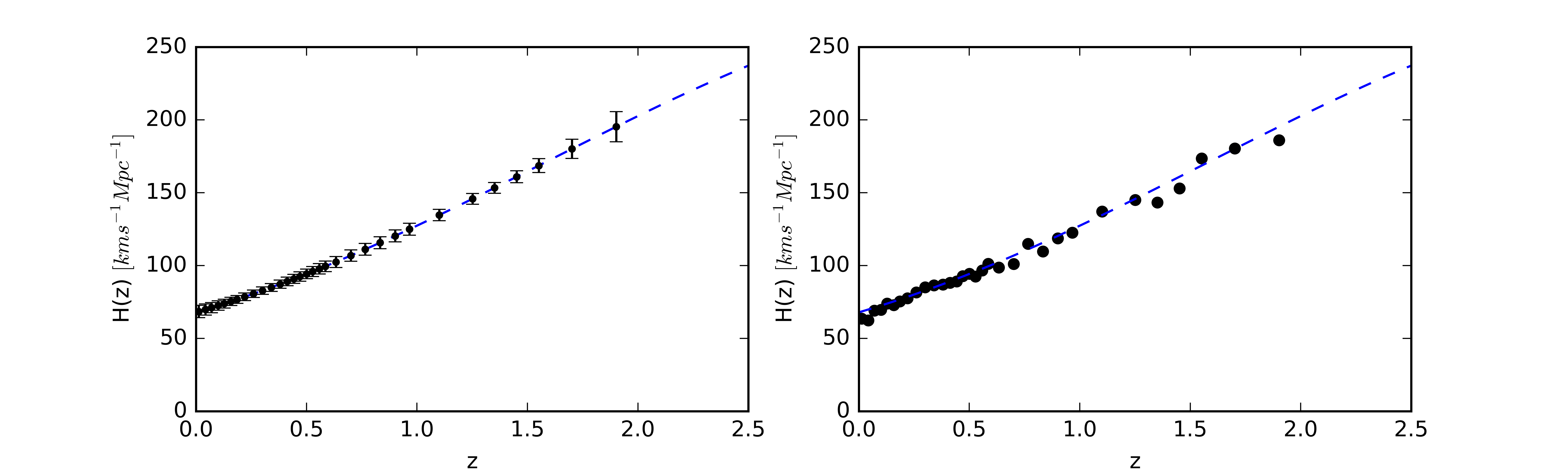}% Here is how to import EPS art
\caption{\label{fig:gaussian simulated h} %gaussian simulated h
Left: The reconstructed Hubble parameter at redshift $z_{sim}$ with 2$\sigma$ error bars. Right: Simulated $H_{sim}(z_{sim})$ in Eq. (\ref{eq:15}). The blue dashed line represents the mean value of the reconstructed $H(z)$.}
\end{figure}

In the first step, we reconstruct the distribution of $H_m(z_{sim})$ using the Gaussian Process with the observational datasets. The Gaussian Process reconstructs the distribution of $H(z)$, and we generate $H_m(z_{sim})$ using the Gaussian distribution $N(\sigma(z_{sim}),\epsilon(z_{sim}))$. $\sigma(z_{sim})$ is the mean of the distribution represented by the black points in the blue line in the left panel of Fig. \ref{fig:gaussian simulated h}, and $\epsilon(z_{sim})$ is shown in the left panel of Fig. \ref{fig:gaussian simulated h} with error bars within the 95$\%$ confidence interval. We select $\epsilon(z_{sim})$ within this interval. The simulated $H_m(z_{sim})$ is shown in the right panel of Fig. \ref{fig:gaussian simulated h}.

\begin{figure}[H]
\centering
\includegraphics[width=8cm]{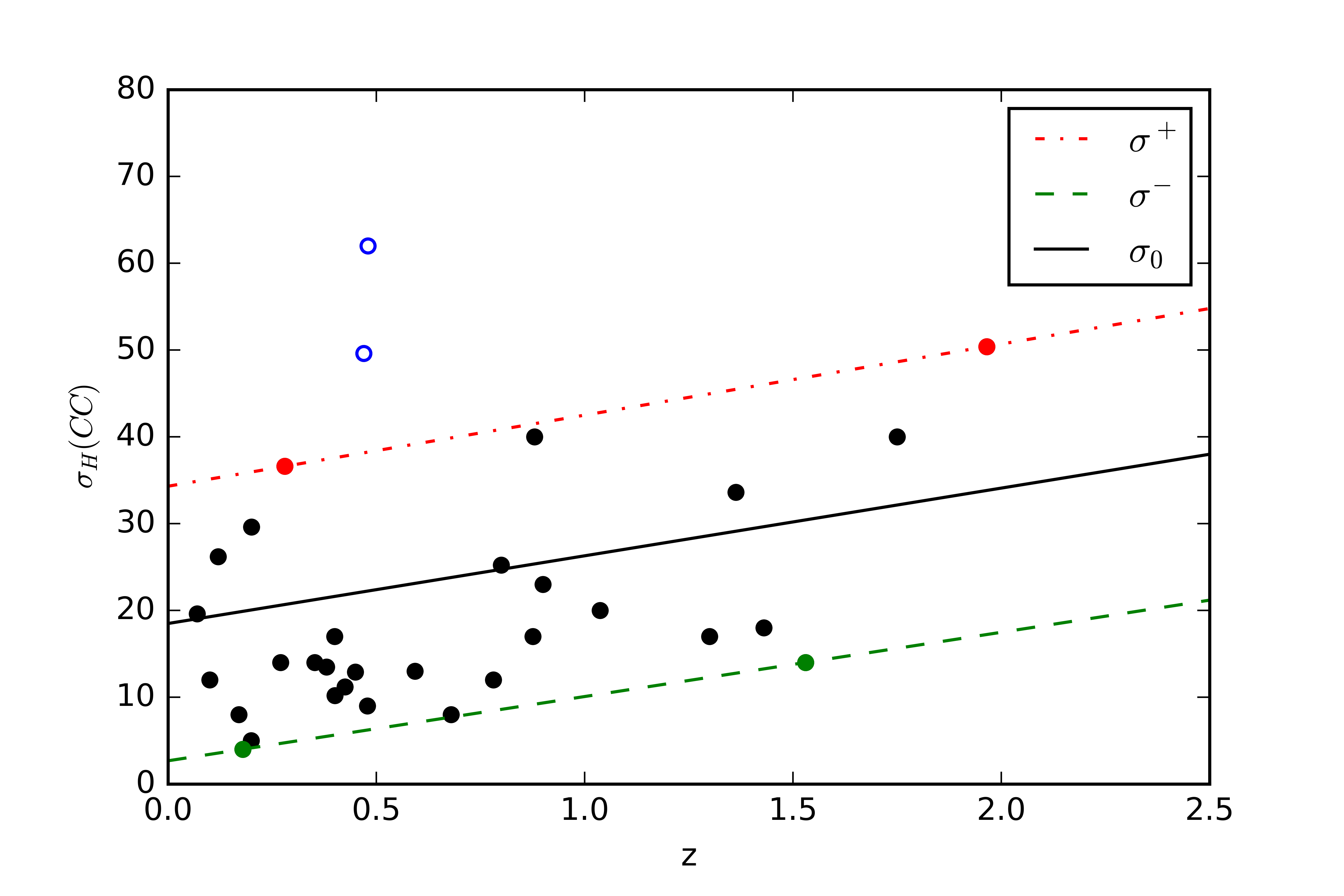}% Here is how to import EPS art
\caption{\label{fig:cc sigma_line} %cc sigma line
1$\sigma$ uncertainty of $H(z)$ in the CC data set. The solid dots and blue circles represent non-outliers and outliers, respectively. The red dash-dotted line represents the upper bound of the data points' uncertainty, denoted as $\sigma^+$, while the green dashed line represents the lower bound of the uncertainty, denoted as $\sigma^-$. The solid black line represents the mean uncertainty $\sigma_0$.}
\end{figure}

\begin{figure}[H]
\centering
\includegraphics[width=8cm]{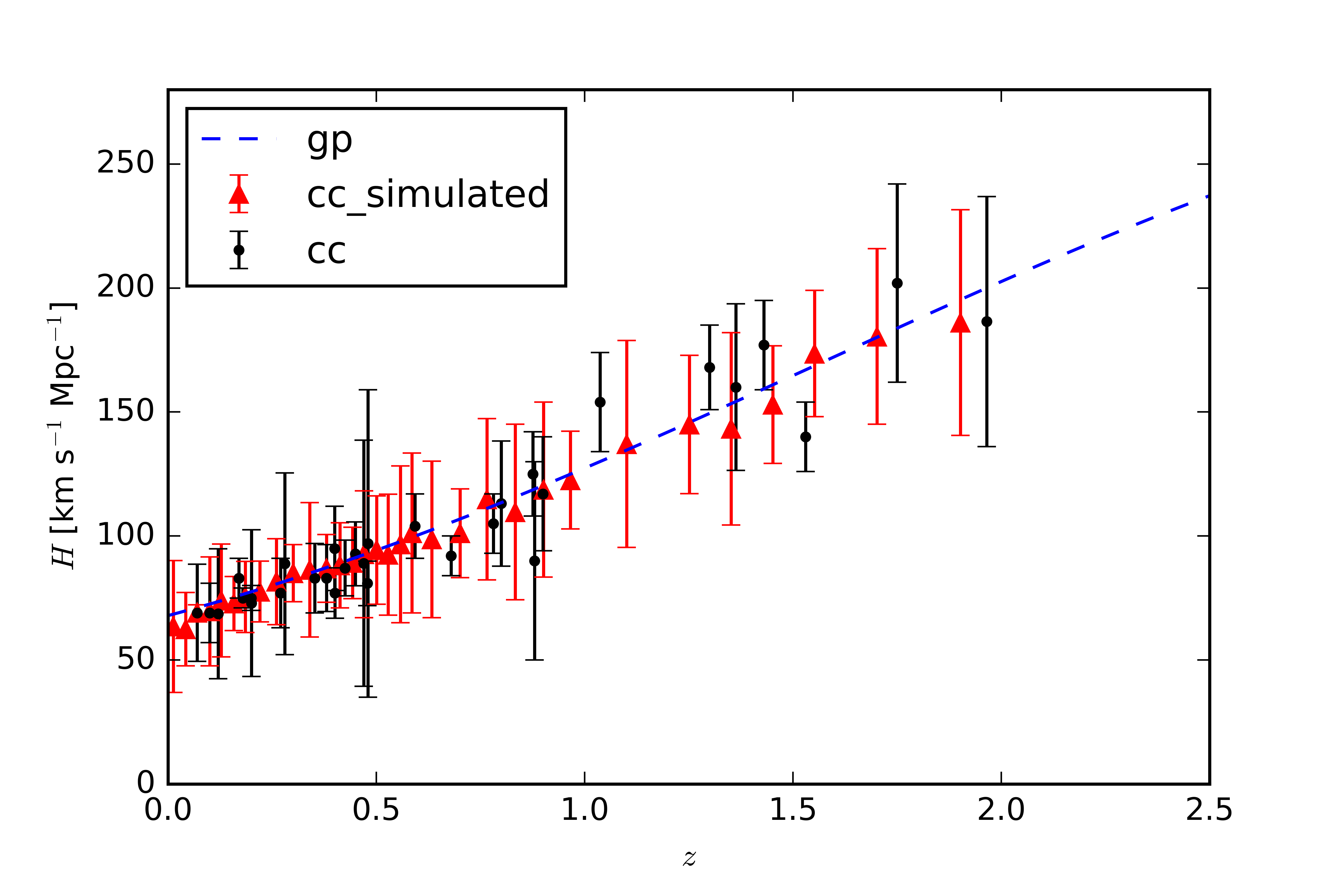}% Here is how to import EPS art
\caption{\label{fig:cc_simulate_z_h_sigma}% cc simulate z h sigma
Simulated $H_{sim}(z_{sim})$ data set based on CC data set using our method. The CC data set is also shown for comparison. The blue dashed line denotes the mean value of reconstructed $H(z)$ from CC data set.}
\end{figure}

In step two, we employ the method proposed by \cite{2011ApJ...730...74M} to generate the uncertainty $\sigma_{sim}$. To assess the uncertainty of the $H(z)$ datasets, we plot the z-$\sigma_H$ in Fig. \ref{fig:cc sigma_line}. In the figure, we exclude the outliers in the blue circle and bound the remaining data points with two straight lines: $\sigma^+=8.2z+34.3$ in red and $\sigma^-=7.4z+2.7$ in green. The uncertainty of $H$ at the $z$ point is estimated by a Gaussian distribution $N(\sigma_0(z),\epsilon(z))$, where the mean is the midpoint of the lines $\sigma_0=7.8z+18.5$ and $\epsilon(z)=(\sigma^+-\sigma^-)/4$. We choose $\epsilon(z)$ such that the generated $\sigma$ falls within 95$\%$ of the Gaussian distribution. Using Eq. (\ref{eq:15}), we obtain the simulated Hubble parameter $H_{sim}(z_{sim})$, as shown in Fig. \ref{fig:cc_simulate_z_h_sigma}.

Using the reconstructed method described in Sec. \ref{sec:Reconstruction} with the CC and simulated CC data points, we can reconstruct the dark energy scalar field potential $V(z)$ for the P18 prior and WMAP9y prior. Similarly, we can reconstruct $V(z)$ for the same priors using BAO and simulated BAO data points, as well as using OHD and simulated OHD data points. The improvement in accuracy of $V(z)$ can be quantified by
\begin{equation}
rate(z)=\frac{\sigma(z) - \sigma_{sim}(z)}{\sigma(z)},
\label{eq:16}
\end{equation}
where $\sigma(z)$ is the error in the reconstructed dark energy scalar field potential $V(z)$ using the observed Hubble parameter at redshift $z$, and $\sigma_{sim}(z)$ is the error in $V(z)$ reconstructed using both the observed and simulated $H(z)$ at redshift $z$. The improvement rate varies depending on the simulated $H(z)$ used, with a range of approximately 5$\%$ to 30$\%$. Thus, doubling the number of Hubble parameter $H(z)$ results in an increase in the accuracy of the dark energy scalar field potential by 5$\%$ to 30$\%$.

\section{Discussions and conclusions}
\label{sec:Discussions and conclusions}

In this paper, we reconstruct the dark energy scalar field potential $V(\phi)$ using Gaussian Process. \cite{2022JCAP...11..037J} and \cite{2022arXiv220306767E} also reconstructed the dark energy scalar field potential using Gaussian Process. The details of the reconstructions are different from each other. To reconstruct the scalar field potential, we need both the dataset and prior, as well as a choice of covariance function for the Gaussian Process. We use the updated CC, BAO, and OHD (CC+BAO) datasets and analyze the influence of different datasets on the reconstructed scalar field potential. In contrast, \cite{2022JCAP...11..037J} used the CC and SN1a datasets, while \cite{2022arXiv220306767E} used only the OHD dataset. Regarding priors, we choose P18 and WMAP9y to analyze their influence on the reconstructed scalar field potential, while \cite{2022JCAP...11..037J} used P18 and a large prior, and \cite{2022arXiv220306767E} used three different priors. For the covariance function, both we and \cite{2022JCAP...11..037J} choose the Squared Exponential function, whereas \cite{2022arXiv220306767E} chose two different covariance functions: the Squared Exponential and the Matern (ν = 9/2) covariance functions. Based on the $\chi^2_{red}$ values computed between the reconstructed $V(z)$ and the models chosen in Section \ref{sec:Reconstruction}, we find that the Power Law model tends to perform better overall than the Free Field model.  Additionally, $V(z)$ reconstructed using BAO$\_$P18 dataset$\_$prior demonstrates a better fit to the models compared to CC$\_$P18, CC$\_$WMAP9y, BAO$\_$WMAP9y, and OHD$\_$WMAP9y. Notably, in the case of OHD$\_$P18, the models are excluded at a significance level of 3$\sigma$ in the high-redshift region ($z\gtrsim 1.5$). Our results show that both different datasets and different priors have a significant impact on the reconstructed $V(\phi)$. The study of dark energy is crucial to understanding the acceleration of the universe. Scalar field is a viable solution to explain dark energy, and the dark energy scalar field potential $V(\phi)$ provides an effective approach to study it. In this paper, we update the available CC and BAO datasets and provide detailed procedures to reconstruct $V(\phi)$ via Gaussian Process. While previous studies have also used Gaussian Process to reconstruct $V(\phi)$ \citep{2022JCAP...11..037J,2022arXiv220306767E}, our study is the first to compare the impact of different datasets and priors on the reconstructed $V(\phi)$.

In Sec. \ref{sec:dark energy scalar field potential}, we present a step-by-step derivation of the dark energy scalar field potential $V(\phi)$ and its 1$\sigma$ uncertainty $\sigma_V$. Then we show the reconstructed $V(\phi)$ using CC, BAO, or OHD datasets with either the P18 or WMAP9y priors in Fig. \ref{fig:CC_p18}, \ref{fig:cc_WMAP9y}, \ref{fig:bao_P18}, \ref{fig:bao_WMAP9y}, \ref{fig:OHD_p18_differentcolor}, and \ref{fig:ohd_WMAP9y_differentcolor}. In the future, as more data points become available and more accurate priors for H$_0$, $\Omega_{M0}$, and $\Omega_{k0}$ are obtained, we can enhance the accuracy and reliability of our $V(z)$ reconstructions. These improvements will allow us to rigorously validate the $V(z)$ models. In the following phase of our work, we will validate existing models by comparing them with the reconstructed $V(z)$ data. Furthermore, we will explore the potential for proposing new models that are consistent with our reconstructed $V(z)$. To assess the impact of different datasets and priors on the reconstructed $V(\phi)$, we compare the results. Our analysis reveals that both different datasets and different priors have a great influence on the reconstructed $V(\phi)$, as shown in Fig. \ref{fig:same dataset; different prior} and \ref{fig:same prior; different dataset}. Especially in Fig. \ref{fig:same prior; different dataset}, we compare the impact on the reconstructed $V(z)$ using three different datasets: CC, BAO, and OHD (CC+BAO). This allows us to observe the influence of different individual datasets (CC vs. BAO) and also compare the individual datasets (CC or BAO) with the combined dataset (OHD). In order to measure the improvement in accuracy of $V(\phi)$ resulting from an increased number of $H(z)$ data points, we conduct a simulation of $H(z)$ data. The simulation process consists two parts: $H_{sim}=H_m(z)+\sigma_{sim}(z)$, where $H_m(z)$ is the mean value of the simulated $H_{sim}(z)$ and $\sigma_{sim}(z)$ is the error of the simulated $H_{sim}(z)$. We simulate $H_m(z_{sim})$ to match the density of observed $H(z)$ data points. By doubling the number of $H(z)$ data points, we find that the accuracy of reconstructed $V(\phi)$ increased by around 5$\%$ to 30$\%$.

The accuracy of the reconstructed $V(\phi)$ is significantly affected by the simulation of $H(z)$. In this study, we simulate $H(z)$ using Gaussian Process based on the observed Hubble parameter $H(z)$. The simulated $H(z)$ varies each time, and the different results of the simulation have a considerable impact on the reconstructed $V(\phi)$. The improvement in accuracy rate can vary from 5$\%$ to 30$\%$ due to the different simulated $H(z)$. As discussed in Sec. \ref{sec:Simulation}, the number of data points is important. Therefore, in the future, we will focus on obtaining more $H(z)$ data to enhance the accuracy of the dark energy scalar field potential $V(\phi)$.

\begin{acknowledgments}
We thank Jie-Feng Chen and Yu-Chen Wang for their useful discussions. This work was supported by the National SKA Program of China (2022SKA0110202)，National Science Foundation of China (Grants No. 11929301).
\end{acknowledgments}

\software{Gaussian Processes in Python \citep{2012JCAP...06..036S}}


\bibliography{sample631}{}
\bibliographystyle{aasjournal}



\end{CJK*}
\end{document}